\documentclass{XrU2005}

\usepackage{graphicx,amssymb,natbibspacing}
\title{X-ray High Resolution and Imaging Spectroscopy of Supernova Remnants}
\author{Jacco Vink}
\affil{Astronomical Institute, University Utrecht, PO Box 80000, NL-3508TA,
Utrecht, The Netherlands}

\newcommand{\adspr}{{\it Adv. Space Res.}}

\newcommand{\xmm}{{\it XMM-New\-ton}}
\newcommand{\chandra}{{\it Chandra}}
\newcommand{\einstein}{{\it Einstein}}
\newcommand{\asca}{{\it ASCA}}
\newcommand{\rosat}{{\it ROSAT}}

\newcommand{\sax}{{\it BeppoSAX}}

\newcommand{\integral}{{\it INTEGRAL}}

\newcommand{\kte}{{$kT_{\rm e}$}}
\newcommand{\net}{{$n_{\rm e}t$}}
\newcommand{\kms}{{km\,s$^{-1}$}}

\newcommand\pcc{cm$^{-3}$}

\newcommand\casa{{Cas\,A}}

\newcommand\rcwes{{RCW\,86}}

\newcommand{\tiff}{{$^{44}$Ti}}

\newcommand{\cofs}{{$^{56}$Co}}
\newcommand{\nifs}{{$^{56}$Ni}}

\newcommand{\halpha}{{H$\alpha$}}

\newcommand{\msun}{{M$_\odot$}}

\begin{document}

\keywords{Supernova remnants; shocks; cosmic rays;X-rays}

\maketitle

\begin{abstract}
The launch of Chandra and XMM-Newton has led to important new findings
concerning the X-ray emission from supernova remnants.
These findings are a result of the high spatial resolution with which
imaging spectroscopy is now possible, but also some useful
results have come out of the grating spectrometers of both X-ray observatories,
despite the extended nature of supernova remnants.
The findings discussed here are the evidence for slow equilibration
of electron and ion temperatures near fast supernova remnant shocks,
the magnetic field amplification near remnant shocks due to cosmic
ray acceleration, a result that has come out of studying narrow filaments
of X-ray synchrotron emission, and finally the recent findings concerning
Fe-rich ejecta in Type Ia remnants and the presence of a 
jet/counter jet system in the Type Ib supernova remnant Cas A.
\end{abstract}

\begin{figure}
\centerline{
\includegraphics[angle=-90,width=0.9\columnwidth]{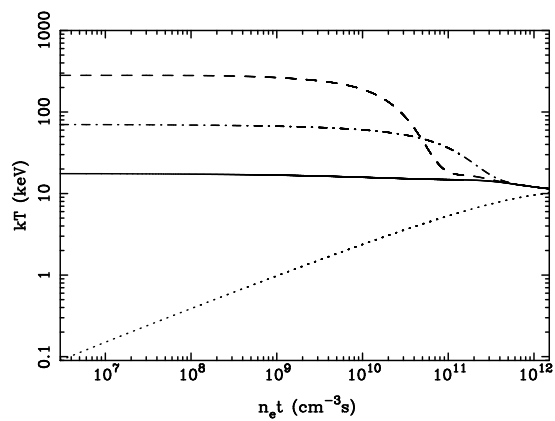}
}
\caption{An illustration of the effect of temperature
non-equilibration at the shock front. Shown is the temperature
of electrons (dotted), protons (solid), helium
(dashed-dotted) and oxygen ions (dashed) as a function of \net,
assuming that heating at the shock front is givenby Eq.~\ref{eq-shocks}.
The oxygen-proton equilibration is
faster than the helium-proton equilibration,
as the cross sections scale linearly with particle mass, 
but quadratically with charge \citep{zeldovich66}.
\label{fig_spitzer}
}
\end{figure}

\begin{figure*}
\centerline{
\parbox{0.68\textwidth}{
\includegraphics[width=0.65\textwidth]{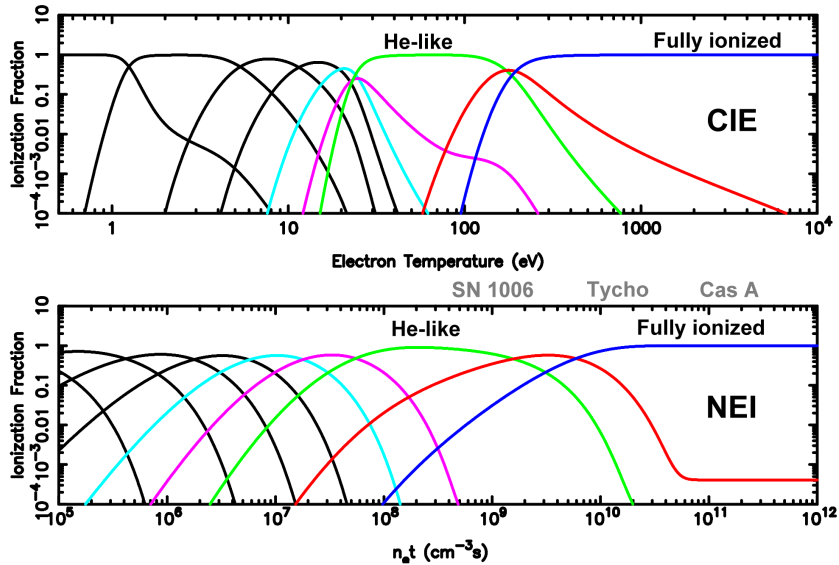}
}
\hskip 0.015\textwidth
\parbox{0.27\textwidth}{
\caption{The effects of non-equilibration ionization (NEI)
illustrated for oxygen.
Both panels look very similar, but the top panel
shows the oxygen ionization fraction as a function of electron
{\em temperature} for collisional ionization equilibrium (CIE),
whereas the bottom panel shows the ionization fraction as function
of \net, and for a fixed  temperature of \kte$=1.5$~keV
\citep[based on][]{shull82}.
Approximate, mean \net\ values for the plasma in the SNRs
\casa, Tycho and SN1006 are indicated.
\label{fig_nei}}
\vskip 7mm}}
\end{figure*}

\section{Introduction}
Supernovae are the most important sources of kinetic energy and chemical
enrichment of galaxies.
By studying supernova remnants (SNRs) we hope to learn about
supernova explosion properties and  chemical yields.
Moreover, because of their energy and large extent, SNRs
are thought to be the principal sources of cosmic
rays of energies up to $\sim 10^{15}$~eV.
Note that the large sizes 
of SNRs are also an important ingredient for their ability
to accelerate cosmic rays, because
astrophysical sources cannot accelerate particles beyond energies 
for which their gyro-radii are larger than the sources themselves.
This means that for cosmic
ray acceleration high magnetic fields and/or large objects sizes are required.

Their large sizes, several parsecs, 
make SNRs also rewarding targets for X-ray imaging spectroscopy.
X-ray imaging spectroscopy made a great leap forward with
the launch of the \chandra\ and \xmm\ satellites.
It is now possible to obtain spectra with a spectral resolution of 
$E/\Delta E \sim 50$ at 6~keV, isolating individual regions 
with an accuracy ranging from 
$\sim$0.5\arcsec\ (\chandra) to $\sim$5\arcsec (\xmm).
Before 1999 \asca\ and \sax\ already provided
imaging spectroscopy, but on arcminutes scales rather than arcseconds
scales. The \einstein\ and \rosat\ imagers on the other hand,
had imaging resolution of  $\sim$5\arcsec, but without any appreciable
spectral resolution.

\chandra\ and \xmm\ also have dispersive high resolution
spectrometers on board, three transmission gratings for \chandra's,
and two Reflective Grating Spectrometers (RGSs) for \xmm.
Due to their dispersive nature, these instruments are not ideal
for spectrometry of extended objects. Nevertheless, for objects of
modest extent, $< 1$~\arcmin, the RGS is still able to obtain
spectra with a resolution of $\lambda/\Delta \lambda > 160$ at 20~\AA.

So what have these instrumental advances brought us, as far as our knowledge
of SNRs is concerned?
The answer is that we have learned substantially about SNR
shocks, concerning both the shock heating process 
and cosmic ray acceleration.
Moreover, imaging spectroscopy has also revealed regions with metal-rich,
pure ejecta plasma, and has provided us with better means to measure
SNR kinematics through X-ray proper motion and Doppler shift studies.

Another important aspect is the study of supernova explosion through
the analysis of kinematics of fresh explosive nucleosynthesis products.
Finally, the high spatial resolution of in particular \chandra\
has enabled the discovery of many young neutron stars inside SNRs,
such as the still enigmatic point source in \casa\ that was discovered in
the first light image of \chandra\ \citep{tananbaum99} (see also
R. Petre these proceedings).
However, in these proceedings I limit myself to three important topics 1) 
collisionless shock physics, 2)  cosmic ray acceleration, and 
3) supernova explosions and nucleosynthesis,

\section{Collisionless shocks physics}

\subsection{Background theory}
\label{sec_nei}
SNR shocks typically move through a medium with 
densities of $n \sim 1$~\pcc. 
At those low densities Coulomb (particle-particle)
interactions are rare, with typical collision times given by 
$1/\tau = 8.8\times10^{-2}/T^{3/2}\ln \Lambda$ \citep{zeldovich66}.
For temperatures of $T = 10^8$~K this gives $\tau \sim 12000$~yr 
( $\ln \Lambda$, the Coulomb is $\sim30$).
This is must longer than ages of many known SNRs.
Nevertheless, we detect X-rays from young SNRs,
which indicates that the plasma got heated despite the long collision times.
This implies that the heating process takes place through long range
collective effects, i.e. the generation of plasma waves. 
This is somewhat analogues to the process of violent relaxation in the
formation  of gravitational systems, such as galaxies and globular clusters.

The fact that supernova remnant shocks are collisionless and are
also sites of cosmic ray acceleration has two important consequences.
First of all, we can no longer assume that different particles species
are in temperature equilibration.
The amount of shock heating as a function of shock speed is obtained
by applying energy, momentum and mass conservation to the gas crossing
the shock front \citep[e.g.][]{mckee80}.
In the extreme case in which particles of different mass do not interact,
the temperature of each plasma component $i$ (i.e.
electrons, protons, other ions) is:
 
\begin{equation}
        kT_{i} = \frac{2(\gamma-1)}{(\gamma+1)^2} m_i v_s^2  = 
	\frac{3}{16} m_i v_s^2,
       \label{eq-shocks}
\end{equation}
where $\gamma$ is the adiabatic index, 
$m_i$, is the particle mass, and $v_s$ is the shock velocity.
For full equilibration this is $kT = 3/16 <m>v_s^2$.
In case full equilibration is not established at the shock front,
Coulomb interactions  will eventually establish equilibration
on a collisional time scales, which is best characterized by
the product of electron density and time \net\ (Fig.~\ref{fig_spitzer}).

\begin{figure*}
{
\parbox{0.5\textwidth}{
\includegraphics[width=0.4\textwidth]{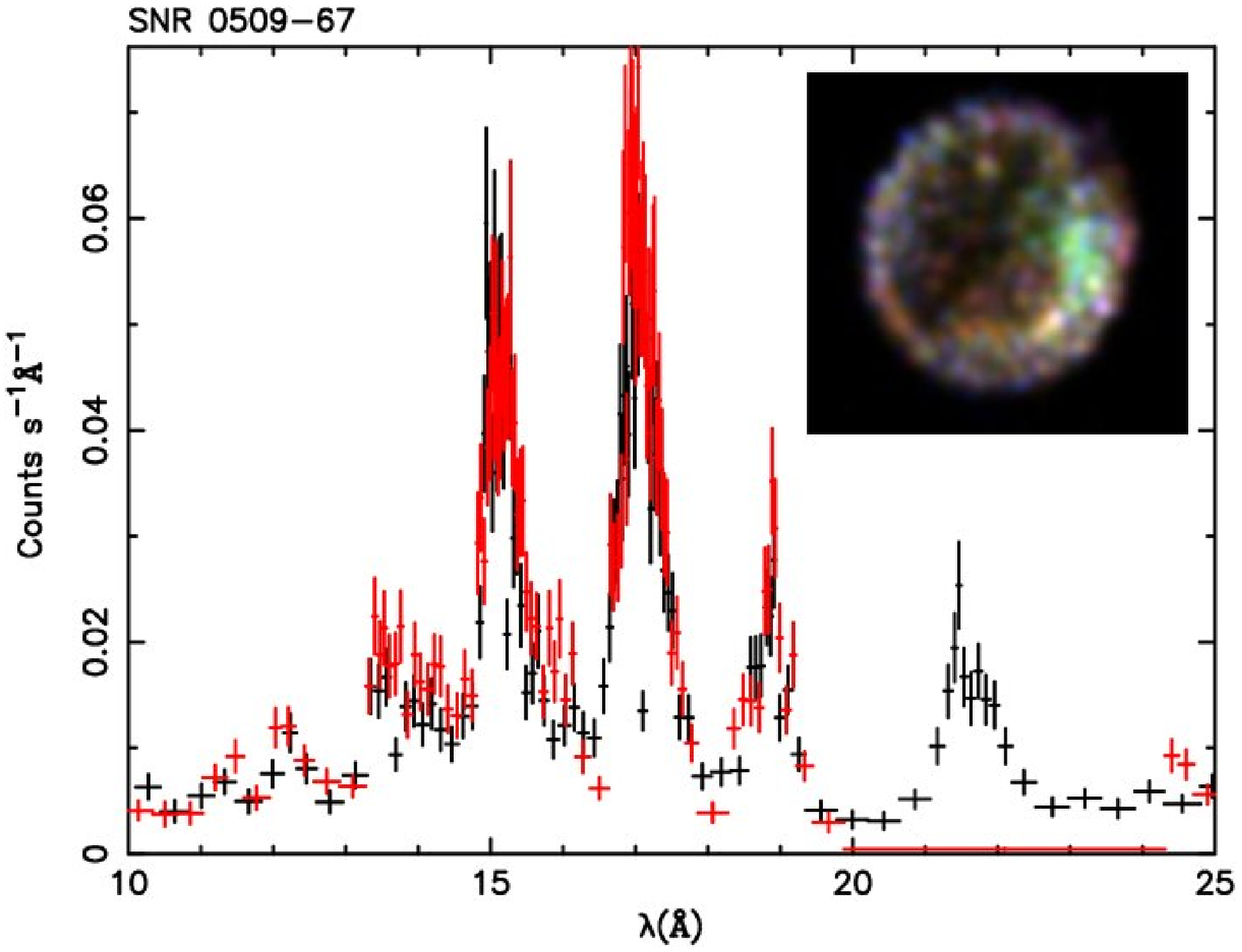}}
\parbox{0.5\textwidth}{
\includegraphics[width=0.4\textwidth]{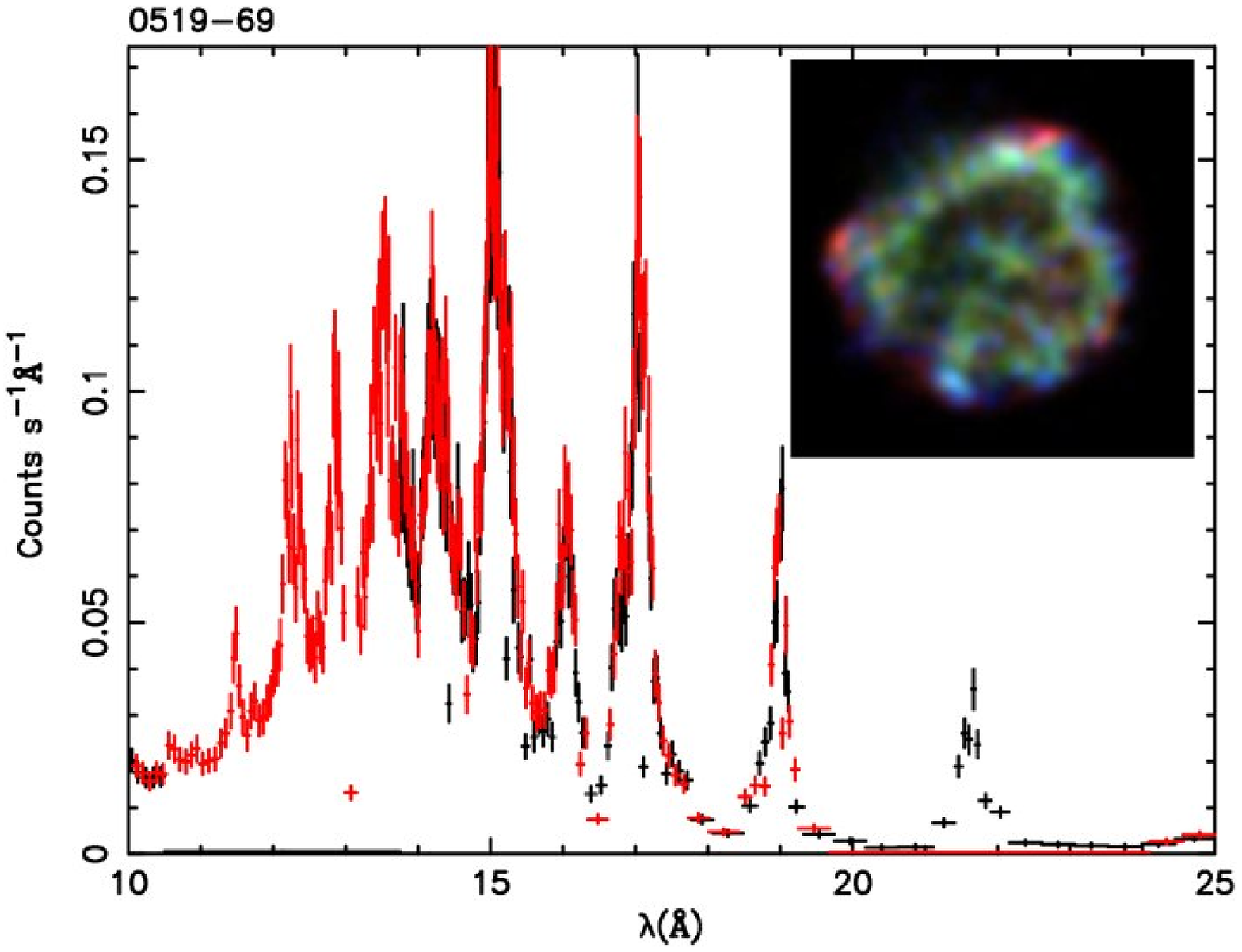}}}
\centerline{
\parbox{0.45\textwidth}{
\includegraphics[angle=-90,width=0.4\textwidth]{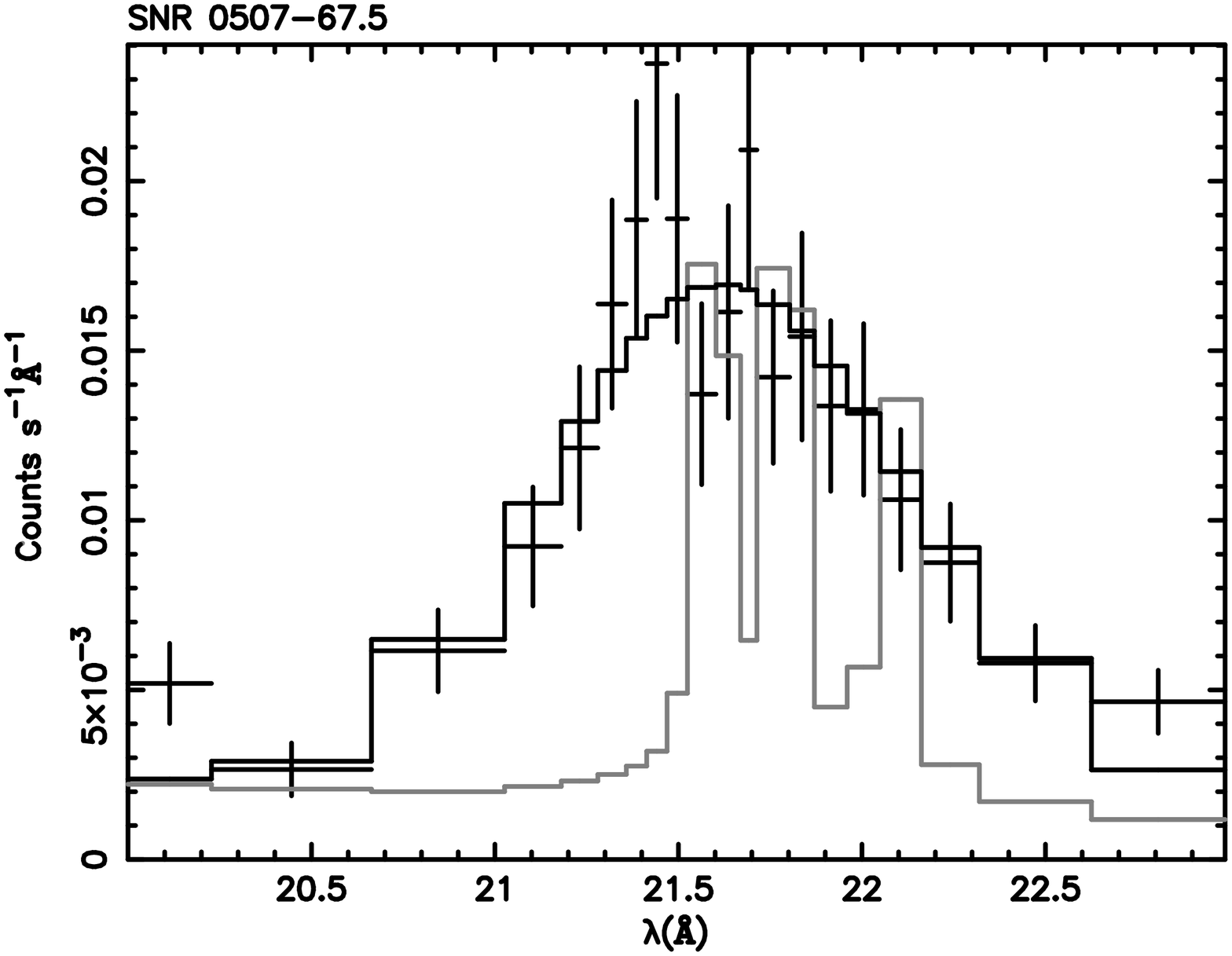}
}
\parbox{0.55\textwidth}{
\caption{Above: Two small SNRs in the Large Magellanic Cloud. 
The spectra are high resolution \xmm\ RGS spectra. The images are
multiband \chandra\ images (see \citet{warren04} for SNR 0509-67.5).
Although the sizes of the two remnants are similar 
SNR 0509-67.5 is the youngest of the two (see text). 
The lines of SNR 0509-67.5 have an extreme velocity broadening
of $\sigma_v \approx 6500$~\kms,
as seen in this close up of the O VII line emission (left,
the intrinsic resolution of the spectrum can be judged
from the gray lines).\label{fig_lmc}
}}}
\end{figure*}

Secondly, Eq.~\ref{eq-shocks} 
assumes that cosmic ray acceleration by the shock 
is energetically not important.
In case the shock also accelerates an appreciable amount of
cosmic rays, the mean plasma temperature may well be lower,
a situation that may have been observed in the supernova remnant
0102.2-7219 \citep{hughes00b}.

When measuring the plasma temperatures by means of X-ray spectroscopy,
one usually measures only the {\em electron} temperature,
as it determines the shape of the bremsstrahlung continuum and it governs
line intensity ratios.
However, because the electron temperature can be lower than 
the average plasma temperature, it is wrong
to infer a shock velocity from measured electron temperatures.
This has been known for quite some time \citep[e.g.][]{itoh77,itoh84},
but until recently it was ignored,
as it was difficult to assess the amount of 
temperature equilibration from the observational data.

Another form of non-equilibration, 
namely non-equilibration of ionization (NEI), has received more attention
over the last two decades, because its signatures could be 
easily discerned in X-ray spectra of SNRs \citep[e.g][]{winkler81,
gronenschild82,jansen88}.

The concept of  NEI is relatively simple \citep{itoh77,mewe80,liedahl99}.
NEI is important in SNRs, because
in plasma that have been relatively recently heated the 
number of electron-ion collisions has been limited.
Collisional ionization equilibrium (CIE) is obtained when
the number of ionizations is compensated 
by the number of recombinations, but for NEI plasmas in SNRs the number
ionization rates still exceed the recombination rate.
The effect of NEI is illustrated in 
Fig.~\ref{fig_nei}. Observationally NEI gives rise to a mismatch between
the electron temperature derived from ratios of line emission
from different ions, 
and the electron temperature derived from the continuum shape, which reflects
the actual electron temperature.
In addition, spectra of NEI plasmas will display lines that are unique
for NEI, and are the result of 
inner shell excitations and ionizations \citep[e.g.][see]{vink04e}.

\begin{figure*}
\centerline{
\parbox{0.35\textwidth}{
\includegraphics[width=0.35\textwidth]{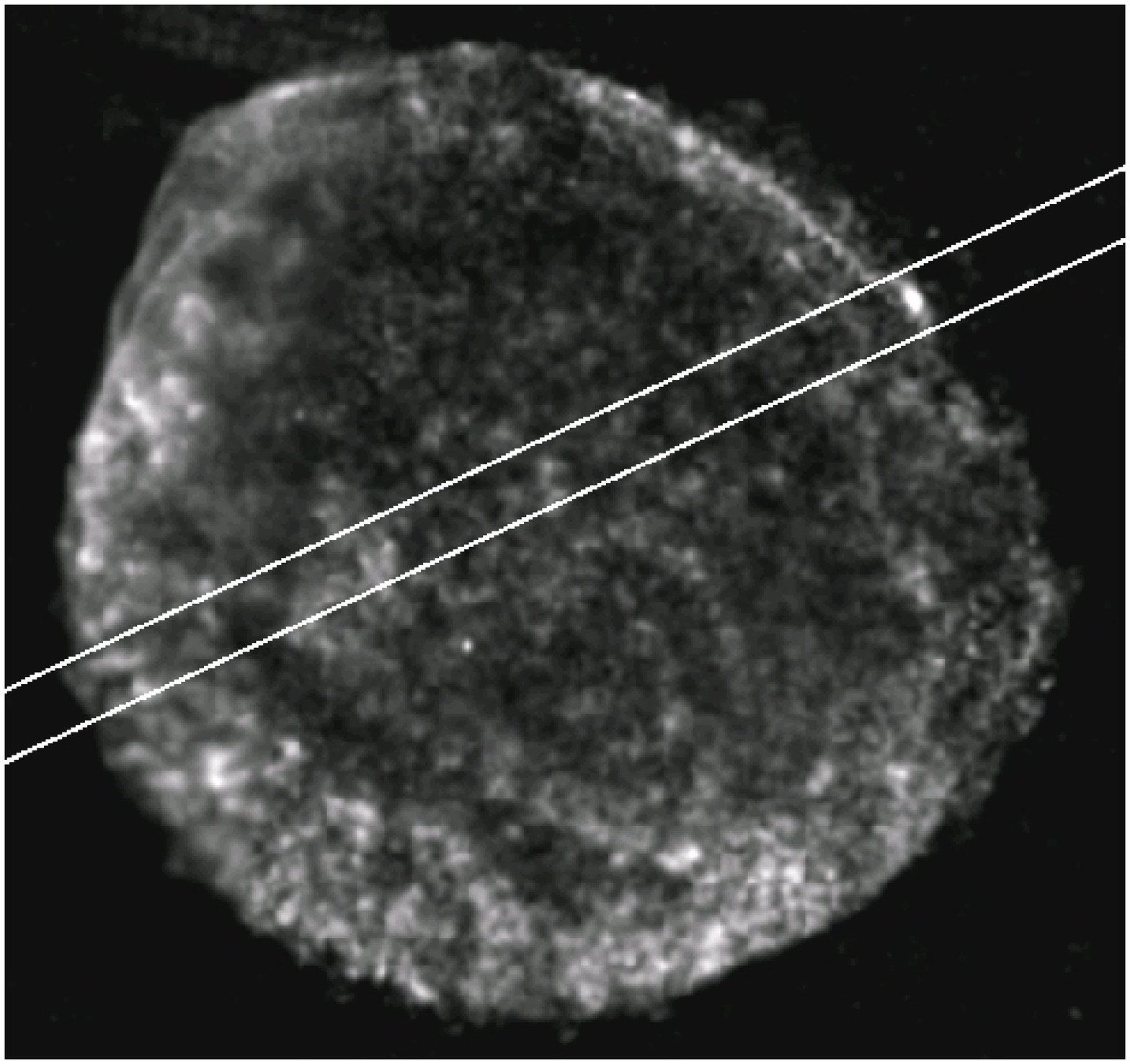}}
\parbox{0.45\textwidth}{
\includegraphics[angle=-90, width=0.45\textwidth]{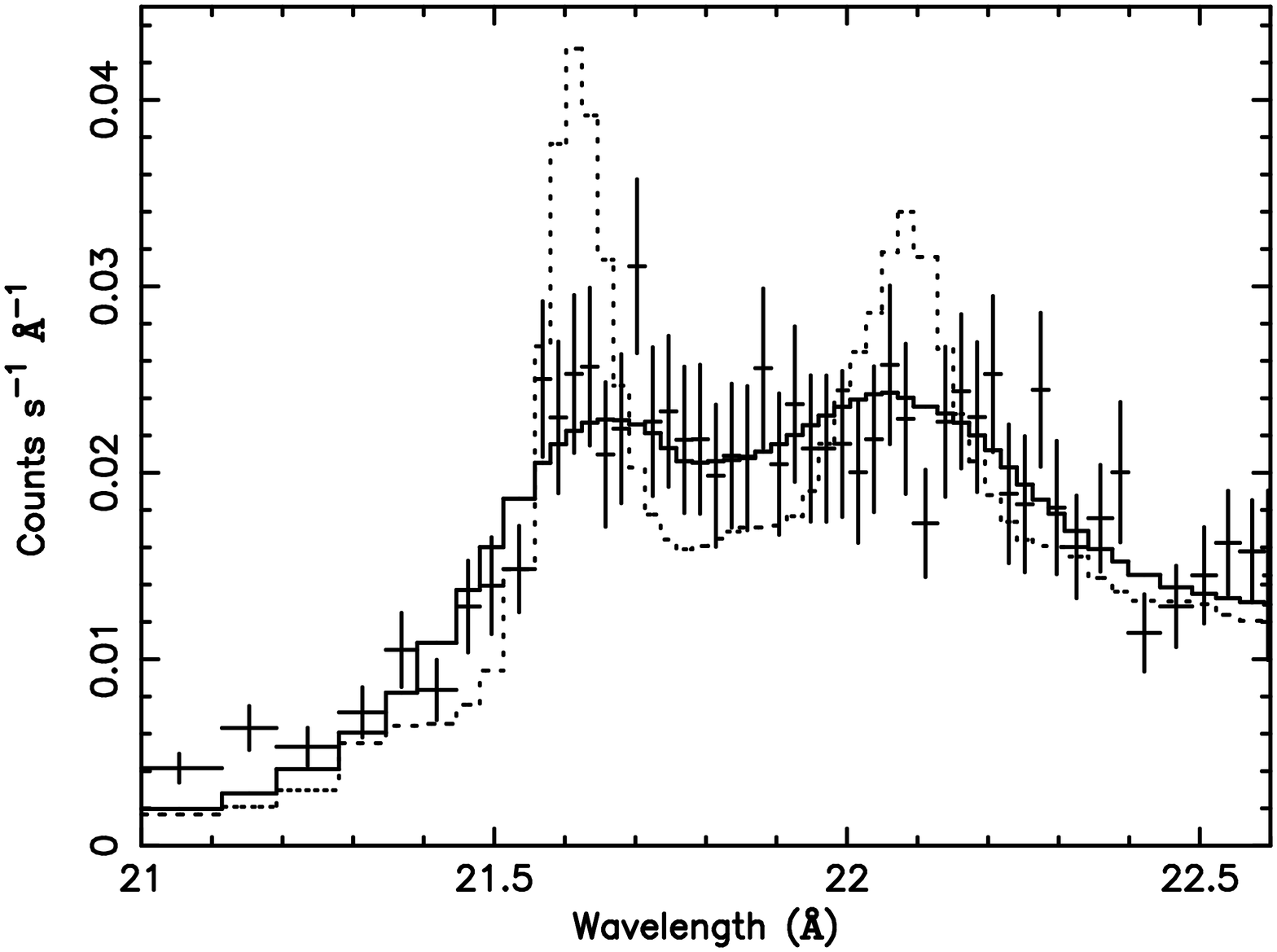}}
}
\caption{On the left: Map of O\,VII emission made from
several \chandra-ACIS observations. The lines indicate
the region observed by the \xmm\ RGS instrument.
The target was the bright knot in the northeast.
Right: Detail of the RGS1 spectrum of the northeastern knot, showing
O\,VII He$\alpha$ line emission.
The dashed line is the best fit model without line broadening,
whereas the solid line shows the model including thermal line 
broadening \citep{vink03b}.
\label{fig_sn1006}}
\end{figure*}

\subsection{High resolution spectra of supernova remnants}
Although the effects of NEI were already known and observed in the eighties,
the \xmm\ RGS instruments make it possible to study it in much more detail.
For extended objects the spectra are blurred due to the spatial extent
of the objects, but the dispersion angle of the RGS is relatively large,
allowing for high resolution spectroscopy of objects that are smaller
than 1~\arcmin.\footnote{For an angular extent of $\Delta \phi$ the
degradation in spectral resolution is
$\Delta \lambda \approx 0.12\Delta \phi$} 
Even for larger objects, up to 5\arcmin, 
one can still obtain useful results with the RGS, especially at long
wavelength, as has been done for the SNRs \casa\ \citep{bleeker01},
Tycho \citep{decourchelle01}, and G292.0+1.8 \citep{vink04f}.

However, the most detailed spectra are obtained for the various bright
remnants in the Magellanic Clouds 
\citep{rasmussen01,behar01,vanderheyden02,vanderheyden03,vanderheyden04}.

Here I illustrate the capabilities of the RGS by showing two spectra
of the Large Magellanic Cloud remnants 0509-67.5 and 0519-69.0.
Both remnants have very similar sizes, resp. 29\arcsec\ and 32\arcsec,
but X-ray spectroscopy reveals, which one is the youngest of the two
(Fig.~\ref{fig_lmc}).
The spectra are dominated by O\,VII and O\,VIII line emission around 18.5~\AA\
and 22~\AA\ and Fe-L shell line emission, but for 0509-67.5 the Fe-L line
emission shows mainly Fe emission at 15.0 and 17.1~\AA, which reveals that
the emission comes from Fe\,XVII (Ne-like Fe). 

SNR 0519-69.0 on the other hand
shows also prominent emission lines at 12.0~\AA, 13.5~\AA, and 14.0~\AA,
an indication that Fe has been ionized up to Fe\,XXI.
This shows that, despite the similarities in size,  SNR 0519-69.0
is in a more evolved state.
Spectral fitting indicates that $\log n_{\rm e}t \sim 10.1$\ for  0509-67.5,
and $\sim 10.4$\ for 0519-69.0.
More evidence that 0509-67.5 is younger comes from the line widths.
Fig.~\ref{fig_lmc} readily shows that the lines of 0509-67.5 are much
broader than those of 0519-69.0. This is not a result of the spatial
extent of the remnants, because SNR 0509-67.5 is the smallest of the two. 
The lines must therefore be broadened by Doppler broadening.
For 0509-67.5 
the O\,VIII Ly-$\alpha$ indicates a Gaussian broadening on top of the spatial
broadening of $\sigma_v = 6500\pm$~\kms. 
For 0519-69.0 this is $\sigma_v = 1700\pm100$~\kms. 
Using the angular size of the remnants, and the distance to the 
Large Magellanic Cloud is 50~kpc, one obtains rough age estimates
of respectively $\sim 500$~yr and $\sim 2000$~yr, assuming free expansion.
For 0519-69.0 free expansion is probably unlikely, using instead the
radius-velocity relation self-similar Sedov solution for a point explosion,
$v_s = \frac{2}{5}r_s/t$, one finds $\sim 800$~yr.

\subsection{An X-ray observation of non-equilibration of temperatures}
Was non-equilibrium ionization a concept known and accepted by X-ray 
astronomers, non-equilibration of temperatures was sometimes
mentioned as an annoying complication for interpreting data of
SNRs, but for the most time it was simply ignored.
Nevertheless, there were indications that non-equilibration is likely
to be important. For example Eq.~\ref{eq-shocks} predicts plasma
temperatures of $kT\sim 30$~keV for shocks velocities of $\sim 5000$~\kms,
applicable to young remnants such as \casa, Tycho and SN1006.
However, no SNR has ever been observed with temperatures
exceeding even 5~keV.

Since 1995, however, more direct measurement of
ion temperatures have indicated the importance of
non-equilibration of temperatures.
These measurements rely on the thermal Doppler broadening
to measure the ion temperature. This has been done
in the optical \citep{ghavamian01,ghavamian03}, 
UV \citep{raymond95,laming96,korreck04}, 
and X-rays \citep{vink03b}. The optical measurements use the fact that
a fraction of the cold neutral  hydrogen undergoes charge exchange
with shock heated protons behind the shock. This results in Doppler
broadened \halpha\ emission.

\begin{figure*}
\centerline{
\parbox{0.4\textwidth}{\vbox{
\includegraphics[width=0.4\textwidth]{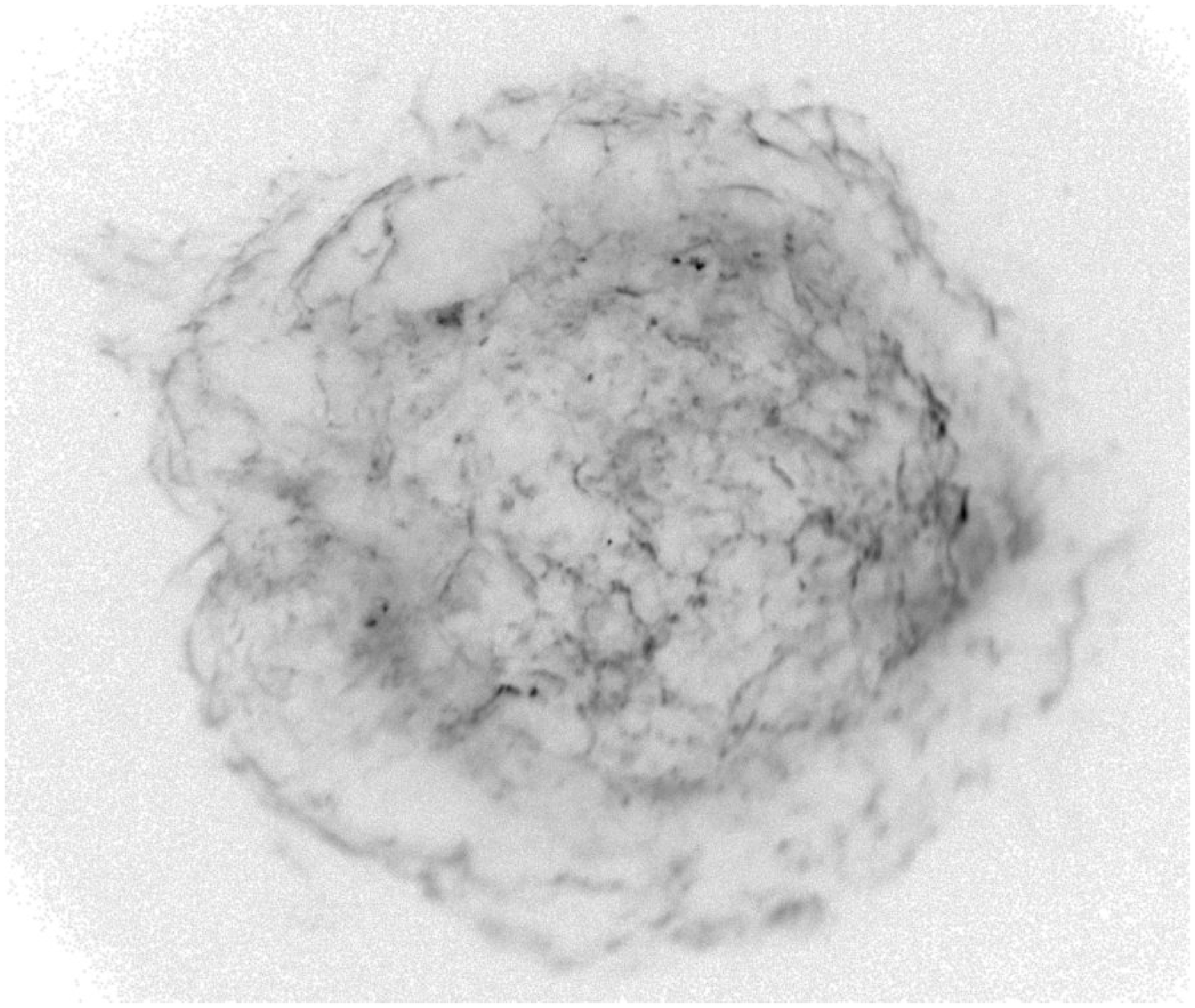}
    \vskip 8mm}
}
\parbox{0.4\textwidth}{\vbox{
     \includegraphics[width=0.4\textwidth]{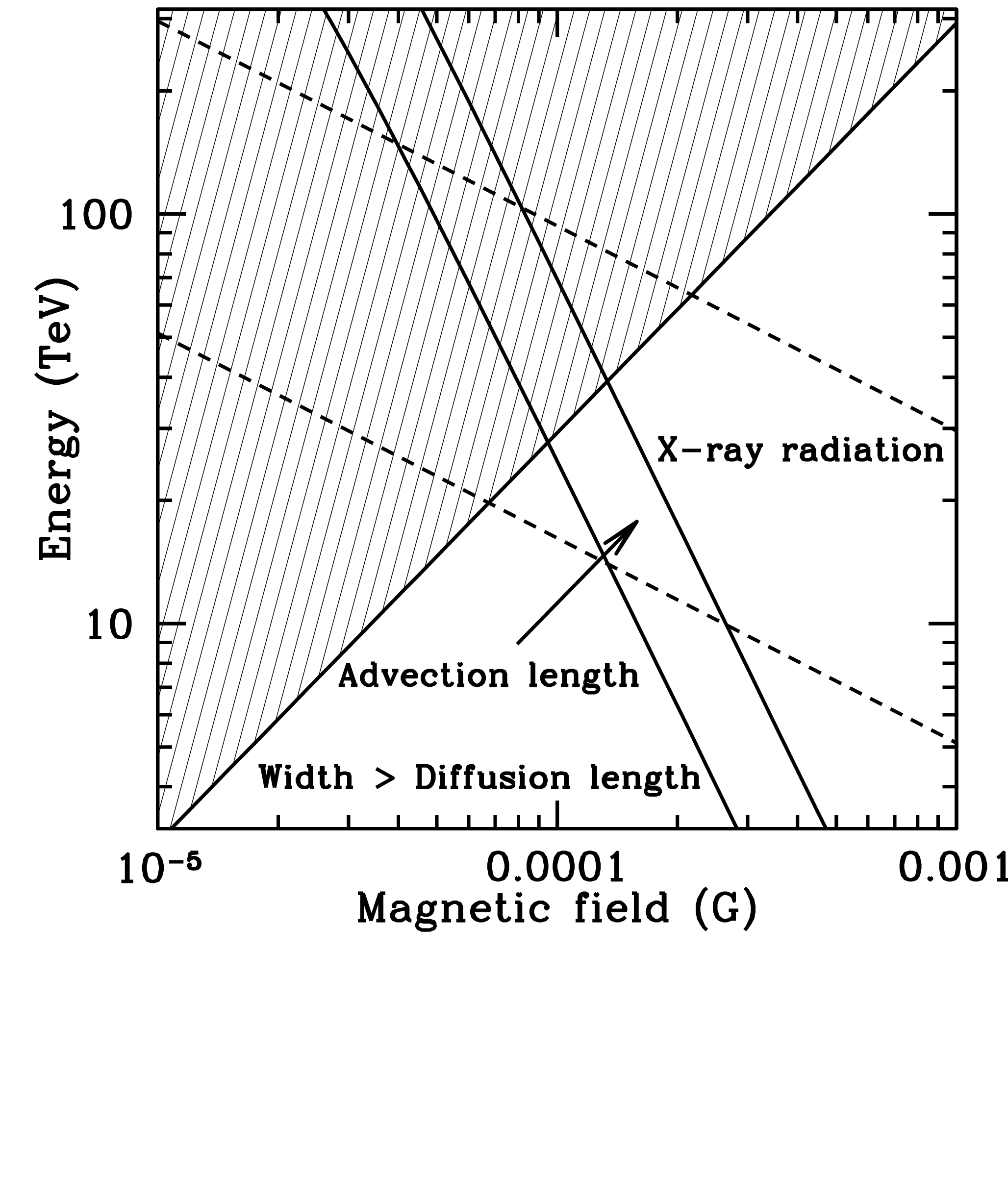}
     \vskip 0mm
  }
}}
  \caption{A deep \chandra\ image of Cas A \citep{hwang04}
in the 4-6 keV continuum band (left).
Note the thin filaments, marking the border of the remnant
(NB the point spread function is not
uniform). The remnant has a radius of about 2.5\arcmin.
Right: Determination of the maximum electron
energy versus magnetic field strength
for the region just downstream of \casa's shock front, as determined
from the thickness of the filaments. The shaded area
is excluded, because the filament width cannot be smaller than the minumum
possible diffusion length \citep[c.f.][]{vink03a}.
\label{bfield}}
\end{figure*}

The first X-ray measurement of the ion temperature was made with the \xmm-RGS.
It may be surprising that the target was a rather extended object: SN1006. 
This SNR has an extent of 30\arcmin. 
There were, however, two reasons to pick SN1006.
First of all the X-ray spectrum shows that it is very far out of ionization
equilibrium $\log$ \net$ = 9.5$. Secondly, in order to isolate
the thermal broadening from bulk motion from the expanding shell one
has to isolate the edge of the remnant. This cannot be done with small
remnants such as those in the Magellanic Clouds. For example, for SNR 0509-67.5
the line broadening is likely to be dominated by the shell expansion.
For a large remnant such as SN1006 one can more easily isolate the outer edge.
What made a  high resolution X-ray spectrum possible despite the large
extent of SN1006, was the fact that at the northeastern edge of
the remnant there is a bright knot with a size of
less than 1~\arcmin (Fig.~\ref{fig_sn1006}).
Nevertheless, the RGS X-ray spectrum of the knot is contaminated by
emission from the inside of SN1006. Luckily the northern part of SN1006
is not very bright, and the X-ray emission from the southern part is attenuated
by the vignetting of the X-ray telescope.

The resulting RGS spectrum of SN1006 consists of lines that have a
sharp rise at the short wavelength side, and a line wing 
at the long wavelength side, caused by
X-ray emission from the inside of the remnant.
Fig.~\ref{fig_sn1006} displays the O\,VII He$\alpha$\ line triplet spectrum
of this knot as observed by the RGS1.
The spectrum can only be satisfactorily fitted with lines broadened with a 
Gaussian distribution with $\sigma_E = 3.4\pm 0.5$~eV, corresponding
to an oxygen temperature of $\sim 500$~keV \citep{vink03b}. 
This seems to be a very high
temperature, but it is in fact what can be expected for shock velocities
of $\sim 4000$~\kms\ in the absence of rapid temperature equilibration
(Eq.~\ref{eq-shocks}).

M. Markevitch showed at this symposium that the
concept of non-equilibration of temperatures is also considered
for shocks in clusters of galaxies. Contrary to SN1006, 
the Mach number $M\sim 3$\ 
shock in the cluster 1E0657-56 is best explained by
rapid equilibration of electron and ion temperatures.
Together with SNR results 
this may contain important information about the physical conditions
that determine the presence or absence of rapid equilibration:

For supernova remnants it has been observed that the fast shocks
of young supernova remnants such as SN1006 and Tycho appear to 
have non-equilibrated plasma's, whereas the slow moving shocks of the
Cygnus Loop ($\sim 300$~\kms) seems to have rapidly equilibrated shocks.
SNRs like Dem L71 and RCW 86 seem to have intermediate equilibration 
properties \citep{rakowski03}.
It is not a priori clear what the physics is behind the different
behavior of fast and slow shocks: 
Is the defining parameter shock speed, Mach number, or Alfven Mach number? 
The observation that the fast shock in 1E0657-56 does seem to equilibrate
rapidly suggests that it is not the shock velocity as such, which is 
$\sim 4500$~\kms\
for 1E0657-56 \citep{markevitch04}. 
Instead it suggest that 
either the Mach number, or the Alfven Mach number is important for the
equilibration process.\footnote{
The Mach number is low in this case because the sound speed is very high.}

\section{Cosmic Ray Acceleration by supernova remnants}
Cosmic rays have been discovered by Victor Hess in 1912 \citep{hess12},
but the question of their origin has still not been satisfactorily
answered. SNRs are the most likely candidate sources for the origin
of cosmic ray energies below $10^{15}$~eV, where there is a break
in the spectrum, usually referred to as the ``knee''. 
It is clear that SNRs can provide the energy to maintain the cosmic
ray energy density in the Galaxy. However, it was for a long time
very doubtful that SNRs were capable of accelerating particles
up to the ``knee'' \citep{lagage83}, 
let alone up to  $10^{18}$~eV, at which energies
the Galaxy becomes ``transparent'' to cosmic rays, and hence, at which
energies there must be a transition from a Galactic origin to an
extra-Galactic origin.

The last five years our understanding of cosmic ray acceleration by
SNRs has changed dramatically, largely due to new theoretical insights,
the coming of age of ground based TeV gamma-ray astronomy,
and last but not least \chandra.

\begin{figure*}
  \begin{center}
\hbox{\hskip 0.03\textwidth
  \begin{minipage}{0.5\textwidth}

		   {  \small    
		     \begin{tabular}[h]{lcccccc}
		       \hline\noalign{\smallskip}
		SNR & Dist & $V_s$& $n_0$ & width & $B_{loss}$ & $B_{diff}$\\
		& kpc  & \kms & \pcc  & \arcsec&  $\mu$G & $\mu$G\\
		\noalign{\smallskip}\hline\noalign{\smallskip}
		Cas A  & 3.4 & 5200 & 3    &0.5& 249  & 299\\
		Kepler & 4.8 & 5300 & 0.35 &1.5& 97   & 113\\
		Tycho  & 2.4 & 4500 & 0.3  &2  &  113 & 165\\
		SN1006 & 2.2 & 4300 & 0.1  &20 &   30 & 39\\
		RCW86  & 2.5 & 3500 & 0.1  &45 &   24 & 14\\
		\hline \\
		     \end{tabular}
		   }
  \end{minipage}

  \begin{minipage}{0.4\textwidth}
    \includegraphics[width=0.9\textwidth]{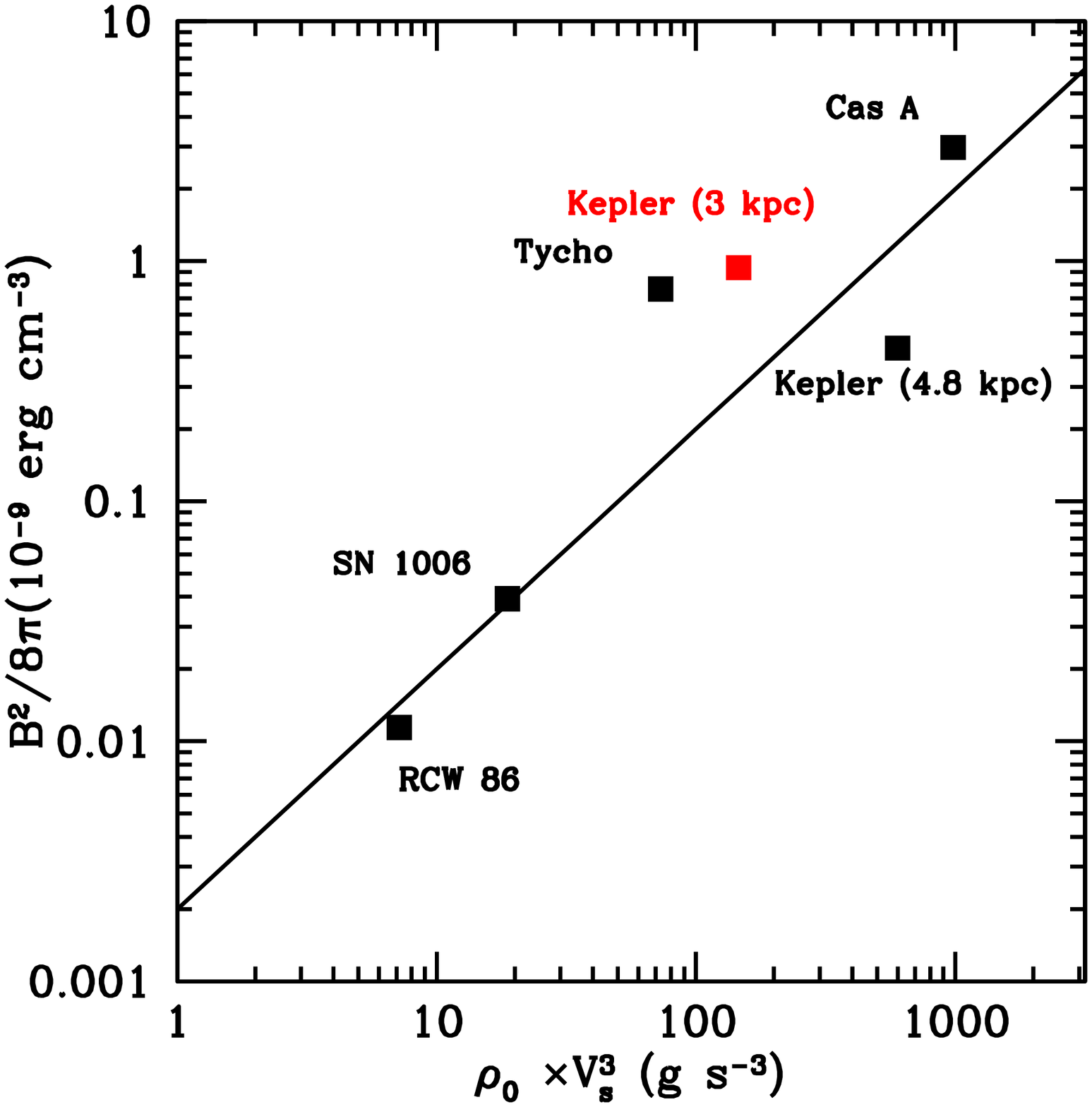}
  \end{minipage}
}
  \end{center}
  \vskip -5mm
    \caption{
Left: Table with magnetic field determinations for various young
supernova remnants \citep[c.f.][]{bamba05,voelk05,ballet05}.
Right: Magnetic field energy density as a function of
$\rho V_s^3$, following the suggestion by \citep{bell04}.
Note that in most cases velocities are based on radio and X-ray
expansion measurements \citep{moffet93,vink98a,hughes99,hughes00c,delaney03},
which introduces a systematic uncertainty for Kepler, for which
no reliable distance estimate exists.
Due to the narrow range in $V_s$\ and large range in densities, the
scaling of $B^2$\ with $\rho$ is more significant than the scaling
with $v^3$. (Vink et al. 2006, in preparation).
      \label{fig_bell}}

\end{figure*}

Cosmic rays are accelerated in supernova remnant shocks by the
first order Fermi process \citep[][for a review]{malkov01}:
Particles are accelerated by repeatedly diffusing across
the shock, at each shock crossing  the particles gains energy
due to the difference in velocity at either side of the shock.
For a high Mach number interstellar shock the compression ratio is 4,
from which follows that the heated plasma behind the shock moves
away from the shock with a velocity $v = 1/4v_s$, with $v_s$ the shock
velocity. The difference in velocity  between the unshocked and shocked
medium is therefore $\Delta v = 3/4v_s$.

The diffusion process itself is a result of elastic scattering
off magnetic field irregularities. 
The fastest possible diffusion, and hence the most efficient cosmic
ray acceleration, is possible when $\delta B/B\sim1$, and is referred to
has Bohm-diffusion. 

The cosmic ray population is dominated by ions, but in X-rays one can only
observe the electron population by means of synchrotron radiation. 
X-ray synchrotron radiation from a shell type SNR is itself a
recent discovery \citep{koyama95}.
The contribution of \chandra\ is that it resolved for the first time
narrow synchrotron filaments near the shock fronts of all young Galactic
SNRs, such as \casa\ and Tycho \citep{gotthelf01a,hwang02}.

The width of these filaments can be interpreted as the result of
advection combined with synchrotron losses:
at TeV energies electron radiation losses are rapid. While diffusing
in the downstream plasma, particles are on average moving away from
the shock front with a velocity $\Delta v$. However, after a time
$\tau_{loss}$\ they are no longer visible in X-rays, because they have lost
a significant fraction of their energy. Hence the observed width
must correspond to $l_{adv} = \Delta v\,\tau_{loss}$\ \citep{vink03a}.
This in itself makes the width a function of $\Delta v $,
particle energy, $E$, and  average magnetic field, $B$, 
since $\tau_{loss} = 635/B^2E$.
In order to separate energy from magnetic field one can use the observed
photon energy $\epsilon$, which depends on $E$\ and $B$\ as
$\epsilon = 7.4 E^2B_\perp$~keV. In Fig.~\ref{bfield} shows graphically
what the allowed range of electron energy and magnetic field is for
\casa: $B \approx 250-500~\mu$G \citep{vink03a,berezhko04a}.
This is much higher than the shock compress mean Galactic magnetic field.
In fact, it turns out that applying this method to all young SNRs gives
relatively high magnetic fields \citep{bamba05,voelk05,ballet05}.

These higher magnetics confirm the hypothesis of Bell and Lucek 
that cosmic ray acceleration gives rise to magnetic field amplification 
due to non-linear plasma wave generation \citep{bell01,bell04}.
This in turns helps to speed up the cosmic ray acceleration process.

\begin{figure}[bt]
\centerline{
\includegraphics[width=0.8\columnwidth]{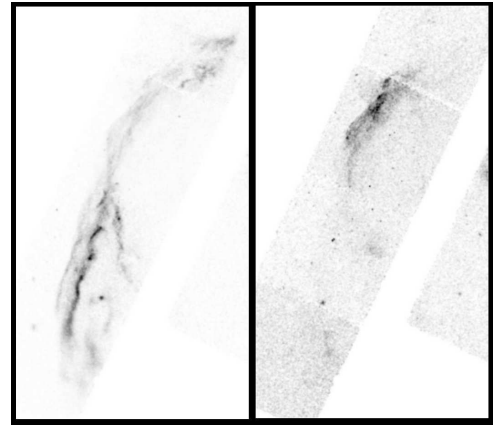}
}
\caption{Two \chandra-ACIS images of the northeastern region
of \rcwes. The left panel shows the energy range from 0.5-1.0 keV,which
is dominated by thermal emission,
whereas the right panel covers the range of 1.9-6.6 keV, dominated by
synchrotron emission. \citep[Vink et al. in preparation, see also][]{vink04d}}
\end{figure}

The magnetic field determinations of different studies differ in that
some \citep{vink03a} have employed the advection length $l_{adv}$, while others
\citep{bamba05,voelk05} have used diffusion length scale, $l_{diff}$.
The diffusion length gives the typical length at which diffusive particle
transport dominates over convection. The diffusion length scale
is given by $l_{diff} = D/\Delta v = \frac{1}{3}c\lambda/\Delta v$, 
with $D$ the diffusion coefficient, and $\lambda$\ the particle mean free path.
For the fastest possible diffusion, Bohm diffusion, $\lambda = E/eB$.
\citet{bamba05} and \citet{voelk05} assumed Bohm diffusion
in order to estimate magnetic field strengths.
This is not a priori correct. However, the observed filament
widths must at least be larger than the $l_{diff}$\ for Bohm diffusion.
In fact when we take this additional constraint into consideration, we
find that the results are consistent with the advection length method 
(Fig~\ref{bfield} and \ref{fig_bell}).

This is not entirely surprising \citep{vink04d}: 
According to shock acceleration
theory the acceleration time to reach an energy $E$, is given by
$\tau_{acc} \approx D/(\Delta v)^2$.\footnote{I have ignored here
some small numerical factor that reflect the fact that the particles
spend time on both sides of the shock front, and
at each side the magnetic field is different due to magnetic field
compression}
For acceleration
we need $\tau_{acc} \leq \tau_{loss}$. 
However, given that $l_{diff} = D/\Delta v$ and 
$l_{adv} = \Delta v\ \tau_{loss}$, we see that 
$\tau_{acc} \leq \tau_{loss}$\ is equivalent
with $l_{diff} \leq l_{adv}$.
So whenever $l_{diff} \approx \l_{adv}$\ acceleration stops and the two
length scales should give approximately the same answer.
However, this is only the case at the very end of the electron
cosmic ray spectrum, where synchrotron losses are important.
This is consistent with X-ray observations, which show steep, loss affected,
synchrotron spectra.

Moreover, the fact that the diffusion length method
{\em assuming Bohm diffusion}
is consistent with the advection length method implies that 
Bohm diffusion does indeed take place 
(see also Markowith, these proceedings, for more details).
This also means that SNRs are more efficient cosmic
ray accelerators than previously thought for two reasons: 1) fast
diffusion applies and 2) magnetic fields are amplified.
In fact assuming Bohm diffusion Cas A must be able to accelerate particles
up to energies of $\sim 2\times10^{15}$~eV for protons and 
$\sim 10^{17}$~eV for iron.
Moreover, the magnetic field amplification is predicted by \citet{bell04}
to scale roughly as $B^2 \propto \rho V_s^3$, which seems indeed
to be the case (Fig.~\ref{fig_bell}). This means that the highest energy
cosmic rays may by those shocks that are fast and have a high density.
These conditions are met by remnants of supernovae with red supergiant
progenitors. These progenitors have a slow and therefore dense wind. 
Moreover, the shock velocity remains high for a long period due to
the $1/r^2$\ density profile. In fact, \casa's progenitor
was probably a red supergiant \citep{vink04a}.

\section{Explosive Nucleosynthesis}
Over the last few years the study of supernovae have become more popular
for two reasons:
One concerns Type Ia  supernovae,
which are likely thermonuclear explosions of white dwarfs.
These supernovae have with great success been used as standard candles 
in cosmology, which has led to the revival of the cosmological constant.
The other reason is that long duration gamma-ray bursts (GRBs) appear to be
associated with special subclass of core collapse supernovae called Type Ib/c
supernovae \citep[e.g.][]{stanek03}.

It is not always possible to associate SNRs with the various
types of supernovae. However, it is clear that all SNRs
with an associated neutron star must be core-collapse supernovae.
Thanks to \chandra\ it is now also clear that a number of young SNRs
in the LMC are likely to be Type Ia SNRs. In all those cases
the SNRs show an increase in Fe abundance toward the center. 
This is exactly what should be expected as Type Ia supernovae are thought to
produce $\sim 0.5$~\msun\ of \nifs\ (which decays into Fe), much more
than the average core collapse supernovae.

It is interesting to use existing SNRs to illustrate
the evolutionary sequence of Type Ia remnants, showing that while
the reverse shock progresses inward into the ejecta, more and more
of Fe gets heated \citep[see][for more sophisticated description of
Type Ia remnant evolution]{badenes05}.
One could start
with 0509-67.5 \citep[Fig.~\ref{fig_lmc}][]{warren04}, 
in which the reverse shock seems to have just reached the Fe layer in the east.
The next one would be 0519-69.0 (Fig.~\ref{fig_lmc}) or alternatively
Tycho's SNR \citep{hwang97,hwang02}, in which more Fe seems to be shocked,
in a shell all around the remnant. The final stage is represented
by DEM L71 \citep{hughes03,vanderheyden03}, 
for which all of $\sim 0.5$~\msun\ of Fe appears to be shocked by the reverse 
shock. In DEM L71 the Fe is no longer in shell, but seems to fill the whole
center of the remnant.

A peculiar case seems to SN1006. The historical supernova is likely to have
been a Type Ia explosion, but both optical/UV absorption \citep{wu97},
and in X-ray emission \citep{vink00a,vink03b,vink04e} there seems to
be a lack of Fe. For the X-ray emission the reason could be that
the reverse shock has not yet reached the Fe-layer. Moreover, the plasma
is so far out of ionization equilibrium (Sect.~\ref{sec_nei}) that even if
Fe is shocked heated it will still have 
an ionization state lower than Fe\,XVII, producing hardly any
emission lines.

The evolution of  remnants of core collapse supernovae are not so easily 
illustrated by a sequence of SNRs. For one thing, the presence of a powerful
pulsar wind nebula does in some cases completely 
dominate the appearance of a SNR.
Moreover, the lack of a well defined sequence may very well reflect the
variety of core collapse progenitors, which is also the reason
why they are unsuitable as standard candles.
Core collapse explosions seem also more turbulent than Type Ia explosions,
so that the initial ordering of stellar layers is not preserved during
the explosion.
The best evidence for that consists of the early appearance of \cofs\ 
radio-active line emission from SN1987A 
\citep[observations are summarized in][]{vink05a}, and the presence of
Fe-rich ejecta {\em outside} the main, Si-rich, 
shell in the southeast of \casa\  \citep{hughes00a,hwang03}. 
In the north the Fe is located inside the Si-rich shell. 
However, this appears to be a projection effect, because the measured 
Doppler velocities of Fe in the north is higher than Si \citep{willingale02}.
It is not clear how much of the Fe in \casa\ is
still unshocked, but some of the shocked Fe must have been ejected with
velocities of up to 7800~\kms.

\begin{figure}
\centerline{
\includegraphics[width=0.85\columnwidth]{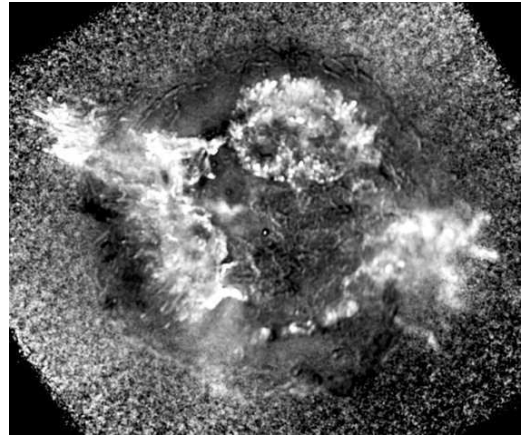}
}
\caption{
Image based on the deep \chandra\ observation of \casa,
which has been processed to bring out the jet/counter jet structure
\citep{vink04a,hwang04}.
(Credit: NASA/CXC/GSFC/U.Hwang et al.)\label{jet}}
\end{figure}

There is no obvious symmetry to the Fe-rich ejecta, so their emergence
is probably related to hydrodynamical instabilities close to the core
of the explosion \citep{kifonidis03}. \casa\ does, however, reveal an
intriguing symmetry when dividing the X-ray map of Si XIII emission
by that of Mg XI emission (Fig.~\ref{jet}). 
It does not only bring out the long known jet in the East, 
but also a counterpart symmetrical situated in the West of the
remnant \citep{vink04a,hwang04}. The spectra of the jet reveal
an apparent absence of Ne and Mg. The dominant elements seem to be 
Si, S, and Ar, but some Fe seems also present.
The emission measure of the jet combined with the average velocity of
the plasma suggest quite a high kinetic energy in the jet, 
$\sim 5\times10^{50}$~erg, about 25\% of the total explosion energy.

The presence of a jet/counter jet system suggests a connection with long 
duration GRBs, which are thought to be the results of beamed
emission from jets. In that case the jets in \casa\ could be the result
of a similar mechanism, although resulting in lower
velocities than for GRB jets. 
However, the point source in \casa\ seems to be neutron star, at odds with the
currently popular collapsar theory for GRBs, for which the a core collapse
into a black hole is needed\citep{macfadyen01}. 
Secondly, GRB jets are thought to
be generated deep inside the star. One would therefore naively expect
the jet material to be dominated by core material, i.e. Fe. 
Still, GRB jets may be electro-magnetic in nature.
Moreover, for the few promising, but still disputed
cases in which line emission from GRBs has been detected, there seems
to be an absence of Fe or Ni emission \citep[e.g.][]{watson03}.
Note in this context that \casa\ has long been thought to be the remnant
of a Type Ib explosion, the same subclass that seems to produce GRBs.

\begin{figure*}
\centerline{
\parbox{0.45\textwidth}{
\includegraphics[width=0.4\textwidth]{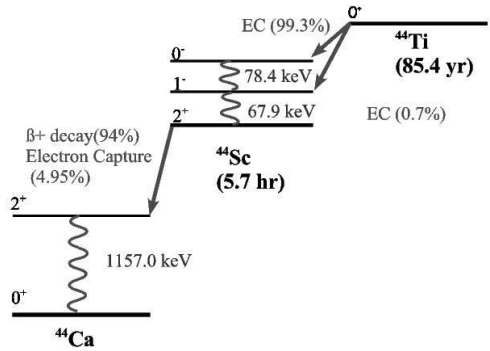}}
\parbox{0.45\textwidth}{
\includegraphics[angle=-90,width=0.4\textwidth]{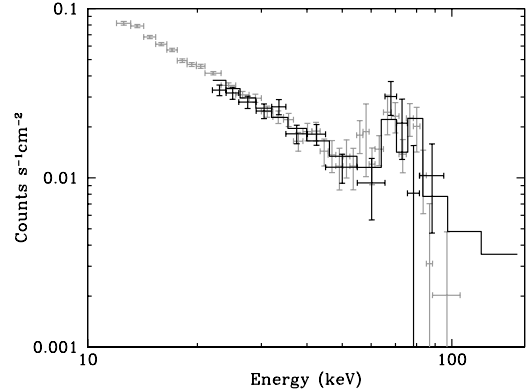}}
}
\caption{
\label{fig_ti44}
Left: The radio active decay scheme of \tiff. Right: hard X-ray spectrum,
as observed by \sax\ (in gray) and \integral, both showing clear
signs of emission around 68~keV and 78~keV due to \tiff\ decay lines
\citep{vink05a}.
}
\end{figure*}

Further exploration  of the 1 million second \chandra\
observation of \casa\  \citep{hwang04}
may reveal more about the nature of its jet/counter jet system.
Moreover, \casa\ is one of two SNRs that is known to contain detectable
amounts of the radio-active element \tiff\ \citep{iyudin94,vink01a}. 
This element is synthesized
deep inside the exploding star, in the same layer as \nifs, but the
\tiff\ yield is very sensitive to the explosion energy and asymmetries.
The observation of \tiff\ in \casa\ is therefore very important for 
studying its explosion properties. This is one of the reasons for observing
\casa\ with INTEGRAL, the first results of which are promising,
but have allowed us to obtain conclusion regarding the kinematics
of \tiff\ \citep[Fig.~\ref{fig_ti44},][]{vink05a}. 
Nevertheless, this is an important goal,
since \tiff\ emission comes both from the shocked and unshocked
parts of the ejecta.

\section{Concluding remarks}
An overview like this is not much more than summing up the conclusions
of many individual studies. So what can I add to that, except
reiterating that the successful launch of \chandra\ and \xmm\ has given
us many new insights into the shock heating, cosmic ray acceleration,
and composition of SNRs.
Many old questions seem to have been (partially) answered, such as the question
of electron-ion temperature equilibration at the shock front (electron-ion
equilibration is not efficient in fast shocks), or the question
whether cosmic rays can be accelerated up to the ``knee'' by supernova
remnant shocks (yes they can).
New questions have been raised by the new observational capabilities,
such as the nature of the jet/counter jet system in \casa. 

In that respect the investigation of SNRs is very similar to
other topics discussed at this meeting.
So let me therefore conclude by thanking the organizers for having invited
me to this very interesting and stimulating meeting.

\end{document}